\documentstyle{article}
\begin{document}
\centerline{\Large \bf Dual Metrics for a Class of Radiative Spacetimes}
\vspace*{0.37truein}
\centerline{D. B{\u a}leanu \footnote{E-mail:
dumitru@cankaya.edu.tr, Institute of Space Science, P.O. Box, MG-23, R
76900, Magurele-Bucharest, Romania }}
\baselineskip=12pt
\centerline{\footnotesize\it Department of Mathematics and Computer Sciences, 
Faculty of Arts and Sciences,}
\baselineskip=10pt
\centerline{\footnotesize\it {\c C}ankaya University, 06531 Ankara, Turkey}
\vspace*{10pt}
\centerline{ S. Ba{\c s}kal
\footnote{E-mail: baskal@newton.physics.metu.edu.tr}}
\baselineskip=12pt
\centerline{\footnotesize\it Physics Department, Middle East Technical
University, Ankara, 06531, Turkey}
%\baselineskip=10pt
%\centerline{\footnotesize\it Ankara, 06531, Turkey}
\vspace*{0.225truein}
\begin{abstract}
\noindent
Second rank non-degenerate Killing tensors for some subclasses 
of spacetimes admitting parallel null one-planes are investigated.
Lichn\'erowicz radiation conditions are imposed to 
provide a physical meaning to spacetimes whose metrics are described 
through their associated second rank Killing tensors.  Conditions under 
which the dual spacetimes retain the same physical properties are 
presented.
\end{abstract}
\vspace*{-0.5pt}
\section{Introduction}   
\vspace*{-0.5pt}
\noindent
Killing tensors are known to be mathematical generalizations of Killing
vector fields, although their defining symmetries are essentially 
different~\cite{ksm}~\cite{rosq89}.
Recently, the reciprocal relations between spaces admitting non-degenerate 
Killing tensors and the spaces whose metrics are specified through those 
Killing tensors have been investigated, together with their generalizations
to Grassman variables~\cite{holten96}~\cite{db01}.
Despite the fact that the geometrical interpretation of the Killing
metrics leads to the notion of geometrical duality, the physical 
signification of those metrics have not been fully understood.
The non-degenerate Killing tensors corresponding to some very well known 
spacetimes, including the Kerr-Newman~\cite{carter68} and the Taub-NUT
metrics~\cite{holten96}, have non-vanishing Einstein tensors, the sources of 
which have neither been identified nor received any interpretation, and 
is still one of the main issues to be clarified.
Mostly, the non-degenerate Killing tensors were constructed from
Killing-Yano\cite{holten95} tensors, but they are not the only solution 
of the Killing tensor equations on those manifolds~\cite{ksm}.
Investigation of the non-degenerate Killing tensors corresponding to the 
Euclidean flat space suggests that there are in fact a class of 
tensors, that are not derived from Killing-Yano tensors~\cite{db01}.  
In that context, the St{\"a}ckel systems of three-dimensional separable 
coordinates were recently investigated~\cite{hinter99}, 
but a four-dimensional and 
non-diagonal example is as yet missing.  Furthermore, finding a class of metrics 
admitting non-degenerate Killing tensors is not an easy task, because the 
condition of non-degeneracy is imposed by hand and has no connection with the 
symmetries of the equations.

The simplest way to attack these problems is to find an appropriate class 
of metrics such that their Killing tensors have the same form as the metric 
itself.  One of those classes can be chosen from spacetimes admitting 
parallel fields of null 1-planes (PN1P), which are not only interesting 
in their own right, but also are appropriate candidates to represent 
radiation with their null vector field indicating the direction of propagation.
The most general form of the metric for fields of parallel m-planes
has been given by Walker~\cite{walker50}.  In four dimensions, for each field 
of non-null and null 1-plane, it is possible to associate a massive and 
a massless particle, respectively, with their four momentum vector to be 
the basis of the 1-planes~\cite{oktem73}.  A very well known subclass of 
Walker's metric is due to Kundt~\cite{kundt61} which has served as a background 
metric for diverse purposes~\cite{podolsky98}~\cite{sb99}.

In this paper, we will classify metrics with PN1P admitting non-degenerate 
Killing tensors.  We will investigate under what conditions, both the initial 
and its dual spacetimes have radiative properties, in an attempt to
contribute to a deeper understanding of dual spacetimes. 

\setcounter{footnote}{0}
\renewcommand{\thefootnote}{\alph{footnote}}

\section{Killing Spaces and Geometric Duality}
\noindent
A second rank Killing tensor is defined through the equation
\begin{equation}
\nabla_{\lambda}K_{\mu \nu}+\nabla_{\mu}K_{\nu \lambda}
+\nabla_{\nu}K_{\lambda \mu}\,=\,0.
\label{kte}
\end{equation}
If this tensor is non-degenerate, then it can be considered as a metric
itself, defining a Killing space~\cite{hinter99} or sometimes more 
specifically a Killing spacetime (KS). It has been shown in detail 
in reference~\cite{holten96} that $K^{\mu\nu}$ and $g^{\mu\nu}$ are 
reciprocally the contravariant components of the Killing tensors with 
respect to each other.  
Then, the second rank non-degenerate tensor $k_{\mu\nu}$,
defined through $K^{\mu \alpha}k_{\alpha \nu}=\delta^{\mu}\,_{\nu}$,
can be viewed as the metric on the ''dual" space.    

The notion of geometric duality extends to that of phase space. 
The constant of motion 
$K=\frac{1}{2}K^{\mu\nu}p_{\mu}p_{\nu}$,
generates symmetry transformations on the phase space linear
in momentum: $\{ x^{\mu},K \}=K^{\mu \nu}p_{\nu}$, and in view of
(\ref{kte}) the Poisson brackets satisfy $\{ H,K \}=0$, 
where $H=\frac{1}{2}g^{\mu\nu}p_{\mu}p_{\nu}$.  Thus, in the phase space
there is a reciprocal model with constant of motion $H$ and the
Hamiltonian $K$. 

The relation between the Christoffel symbols, 
$\hat\Gamma^{\mu}\,_{\alpha \beta}$ of the KS and of the initial manifold 
has been expressed earlier~\cite{db01}.
By writing $\hat\Gamma^{\mu}\,_{\alpha \beta}$ in terms of the
Killing tensor and taking (\ref{kte}) into account we have:
\begin{equation}
\hat \Gamma^{\mu}\,_{\alpha \beta} = \Gamma^{\mu}\,_{\alpha \beta}
-{\cal K}^{\mu\delta}\nabla_{\delta}K_{\alpha\beta},
\end{equation}
where ${\cal K}^{\mu \alpha}K_{\alpha \nu}=\delta^{\mu}\,_{\nu}$.

\section{Parallel Null One-Planes and Lichn{\'e}rowicz \\ 
Radiation Conditions}
\noindent
A parallel field of null 1-plane in spacetime consists in a recurrent field
of null vectors. If $l_{\mu}$ is a basis for the plane we have:
\begin{equation}
\nabla_{\nu} \, l_{\mu} = \kappa_{\nu} l_{\mu},
\qquad 
l_{\mu}l^{\mu}=0, \quad l_{\mu} \ne 0
\label{pn1p}
\end{equation}
where $\kappa_{\mu}$ is the recurrence vector of the plane. 
It can be seen that such a vector field is geodesic and non-rotating.
If the PN1P is strictly parallel then the recurrence vector vanishes
identically.

One criteria for the existence of radiation in spacetime has been 
proposed by Lichn\'erowicz, relies on an analogy with electromagnetism,
and is based on the solution of Cauchy's problem for Einstein-Maxwell
equations~\cite{lich58}.  In brief, the spacetime metric is subject to the 
conditions:
\begin{eqnarray}
&& l^{\mu}R_{\mu \nu \, \alpha \beta}=0,  \label{lrc1} \\
&& l_{[\mu}R_{\nu\sigma ]\,\alpha \beta}=0
\label{lrc2} 
\end{eqnarray}
with $l_{\mu} \ne 0$ and $R_{\mu \nu \, \alpha \beta} \ne 0$.
He also proved that the trajectories of $l^{\mu}$ are null geodesics
if $R_{\mu \nu \alpha \beta} \neq 0$~\cite{lich60}.  
The applicability of Lichn\'erowicz radiation conditions (LRC) 
to some alternative approaches to gravity has been another subject 
of interest~\cite{sb99}.

From (\ref{pn1p}) and the Ricci identity we have
\begin{equation}
l^{\nu}R_{\mu \nu \, \alpha \beta}=l_{\mu}f_{\alpha \beta},
\end{equation}
where
\begin{equation}
f_{\alpha \beta}=\partial_{\alpha}\kappa_{\beta} 
- \partial_{\beta}\kappa_{\alpha}.
\end{equation}
From above, it is seen that when $f_{\alpha\beta}=0$,
one of the radiation conditions due to Lichn\'erowicz is satisfied.

The canonical form for PN1P has been given by Walker as:~\cite{walker50} 
\begin{equation}
ds^{2}=2\,dv\,du\,+\,A'\, dx^{2}\,+\,2 D' \, dx \,dy\,
+\,2E'\,dx \,du\,
+\,B'\,dy^{2}\,+\,2F'\,dy\,du+H'\,du^{2} \, ,
\label{met1}
\end{equation}
where $H'$ depends on $v,x,y,u$ and  $A',B',D',E',F'$ depend only on 
$x,y,u$, with $(A'B'-D'^{2}) > \,0$.  The plane is strictly parallel when 
$H'$ is also independent of $v$.  
It is apparent that the existence of a geodesic null vector field 
is crucial for the description of a radiative spacetime. Therefore, 
Walker's metric is an appropriate candidate if we are to seek spacetimes 
having radiative properties.  

From (\ref{pn1p}) one can observe that the principal null vector
$l_{\mu}=\delta_{\mu}\,^{4}=\partial_{\mu} u$, is hypersurface-orthogonal and
the recurrence vector for the PN1P is $\kappa_{\mu}=-\Gamma^{4}\,_{\mu 4}$.
If both the initial and its KS admit PN1P with the same principal 
null vector then the relation between their recurrence vectors 
can be expressed as
\begin{equation}
\hat\kappa_{\mu}=\kappa_{\mu}+{\cal K}^{4 \delta}\nabla_{\delta}K_{\mu 4}.
\end{equation}

\section{Subclasses of Parallel Null 1-Planes}
\noindent
It has been shown that the canonical form of Walker's metric can be 
brought into a simpler form, by an appropriate choice of the coordinate 
system where $E=F=0$~\cite{akcag78}.  Then the resulting form can always be 
diagonalized within the metric functions $A',B'$ and $D'$. As such the 
simplified form becomes:
\begin{equation}
ds^{2}=2\,dv\,du + A(x,y,u)\, dx^{2} + B(x,y,u)\,dy^{2} + 
H(v,x,y,u)\,du^{2}
\label{mm}
\end{equation}
where $A(x,y,u),B(x,y,u)$ and $H(v,x,y,u)$ are functions of their arguments.

\subsection{PN1P satisfying LRC}
\noindent
We are looking for a subclass of the above metric satisfying LRC.  
Condition (\ref{lrc1}) suggests that all $R_{1\nu\,\alpha\beta}$ vanish
for all $\nu, \alpha, \beta$.  
This in return yields the metric function $H$ in (\ref{mm}) to be linear 
in $v$.
With this there are left only six non-vanishing components of the 
Riemann tensor which are: $R_{23\,23},\,R_{23\,24},\, R_{23\,34},\, \\
R_{24\,24},\,R_{24\,34},\,R_{34\,34}$.  Furthermore, condition (\ref{lrc2}) 
imposes the following restriction on the Riemann tensor: 
\begin{equation}
R_{23\,23} = R_{23\,24} = R_{23\,34}=0,
\label{rizero}
\end{equation}
which are second order non-linear 
coupled equations with respect to the metric functions 
$A(x,y,u)$ and $B(x,y,u)$.  
Analyzing (\ref{rizero}),
two possible solutions can be distinguished:\\
{\bf i.}~~ $A(x,u)$ is independent of $y$ and $B(y,u)$ is independent 
of $x$.  \\
{\bf ii.}~~$A(x,y,u)=B(x,y,u)$. \\
The above spacetimes satisfy Lichn\'erowicz radiation conditions.

\subsection{Subclasses of PN1P admitting non-degenerate Killing tensors}
\noindent
The second step is to classify the above metrics whose Killing tensors
are of the same form. Thus their KS will retain the same properties as 
that of the initial ones.  Corresponding to the cases which we have found 
in the preceding section, we have the following classifications:\\
{\bf Case i.}~~The metrics satisfying the first case in above are of the
following form: 
\begin{equation}
\begin{array}{l}
A(x,u)=a(x)\,(s_{2}(u)-1), \qquad
B(y,u)=b(y)\,(s_{3}(u)-1),\\
H(v,x,y,u)  =v\, h(u)+(r_{1}(x)\,q_{1}(u)
+r_{2}(y)\,q_{2}(u))(s_{2}(u)-s_{3}(u))^{-1}\, , 
\end{array}
\label{case1}
\end{equation}
with $s_{2}(u) \neq s_{3}(u) \neq 1 $.
Here, once and for all, the functions specifying the metric and the Killing
tensor are arbitrary functions of their arguments, unless restrictions 
are explicitly stated.  Solving equation (\ref{kte})
the non-vanishing components of the corresponding Killing tensor are found 
to be
\begin{equation}
\begin{array}{l}
K_{14} =1,\qquad  K_{22}  = A(x,u)\, s_{2}(u) ,\qquad 
K_{33} = B(y,u)\, s_{3}(u), \\[1mm]
K_{44} = H(v,x,y,u) + \left[ r_{1}(x)q_{1}(u)\,(1-s_{2}(u))
+r_{2}(y)q_{2}(u)\,(1-s_{3}(u))\right] \\[1mm]
\qquad \quad (s_{2}(u)-s_{3}(u))^{-1}\, . 
\end{array}
\end{equation}
Here, the functions $q_{1}(u)$ and $q_{2}(u)$ are subject to the following 
equations:
\begin{equation}
\begin{array}{c} 
f_{1}(u)\, q_{1}(u) - g_{1}(u)\, q_{1}(u)_{,u} =0 ,\\[1mm]
f_{2}(u) \, q_{2}(u) + g_{2}(u)\, q_{2}(u)_{,u} =0
\end{array}
\end{equation}
with
\begin{equation}
\begin{array}{l}
f_{1}(u) = s_{2}(u)_{,u}\, (s_{3}(u) - 1)-s_{3}(u)_{,u}(s_{2}(u)-1) 
   - h(u)\,(s_{2}(u)-1)(s_{2}(u) - s_{3}(u)),\\[1mm]
f_{2}(u)  = s_{2}(u)_{,u}(s_{3}(u)-1) - s_{3}(u)_{,u} (s_{2}(u) -1) 
   - h(u)\,(s_{3}(u)-1)(s_{2}(u) - s_{3}(u)), \\[1mm]
g_{1}(u) = (s_{2}(u) - 1)(s_{3}(u) - s_{2}(u)), \\[1mm]
g_{2}(u) = (s_{3}(u) - 1)(s_{3}(u) - s_{2}(u)),
\end{array}
\end{equation}
where the comma denotes partial differentiation.
We found the class of dual metrics corresponding to 
(\ref{case1}) as:
\begin{equation}
\begin{array}{l}
k_{14} =1,\qquad  k_{22}  = A(x,u) s_{2}(u)^{-1} ,\qquad 
k_{33} = B(y,u) s_{3}(u)^{-1}, \\[1mm]
k_{44} = v\,h_{1}(u)+\left[ r_{1}(x)q_{1}(u)s_{2}(u)
+r_{2}(y)q_{2}(u)s_{3}(u)\right](s_{2}(u)-s_{3}(u))^{-1} \, ,
 \end{array}
\end{equation}
where $s_{2} \neq 0,\, s_{3} \neq 0$, 
with further restrictions as in (\ref{case1}).\\   
{\bf Case ii.}~~For the second case we discussed in the previous section,
we have the following metrics:
\begin{equation}
A(x,y,u)=e^{x+y} a(u), \qquad H(v,x,y,u)=v h_{1}(u)+h_{2}(x,y,u).                                                               
\end{equation}
The non-vanishing components of the Killing tensors are now found as:
\begin{equation}
\begin{array}{l}
K_{14}=1,\quad K_{22}=K_{33}=A(x,y.u) (1+a(u)), \\[1mm]
K_{44}=H(v,x,y,u)-a(u) h_{2}(x,y,u) + k_{1}(u),
\end{array}
\end{equation}
where
\begin{equation}
h_{2}(x,y,u)=\frac{e^{P(u)}}{a(u)}
\left[\int e^{P(u)}(h_{1}(u)k_{1}(u)+k_{1}(u)_{,u})du+k_{2}(x,y)\right],
\end{equation}
with $P(u)=\int h_{1}(u)du$.
The corresponding dual metrics become:
\begin{equation}
\begin{array}{l}
k_{14}=1,\qquad k_{22}=k_{33}=A(x,y,u), \\[1mm]
k_{44}= H(v,x,y,u) + a(u)h_{2}(x,y,u) - k_{1}(u) .
\end{array}
\end{equation}

In the following we investigate a very familiar subclass of Walker's metric, 
recognized as Kundt's metric~\cite{kundt61}, to provide an example whose 
Killing tensor is not of the same form of the initial metric but still 
falls into the class of PN1P.  They describe  plane fronted waves with 
parallel rays, admitting a non-expanding shear-free and twist-free null 
geodesic congruence.  This metric is expressed as:
\begin{equation}
ds^{2}=2\,dv\,du+ dx^{2}
+dy^{2}+H(x,y,u)\,du^{2}.
\label{kundt}
\end{equation}   
The metric function $H(x,y,u)$ has either of the following forms:
\begin{equation}
H^{(1)}(x,y,u)=h_{1}(u)-h_{2}(x-y)+x, \, \quad
H^{(2)}(x,y,u)=h_{1}(u)-h_{2}(x-y)+y 
\end{equation}
and admits a Killing tensor in a more general form, whose non-vanishing 
components are
\begin{equation}
\begin{array}{l}
K_{14}=K_{22}=K_{33}=1,\qquad 
K_{24}=K_{34}=u, \\[1mm]
K^{(1)}_{44}=H^{(1)}(x,y,u)-2(x+y)-u^{2}/2 , \\[1mm]
K^{(2)}_{44}=H^{(2)}(x,y,u)-2(x+y)-u^{2}/2.
\end{array}
\end{equation}
The associated dual metrics are found as:
\begin{equation}
\begin{array}{l}
k_{14}=k_{22}=k_{33}=1, \qquad  k_{24}=k_{34}=-u, \\[1mm]
k^{(1)}_{44}=H^{(1)}(x,y,u)+2(x+y)+\frac{5}{2}u^{2},\\[1mm]
k^{(2)}_{44}=H^{(2)}(x,y,u)+2(x+y)+\frac{5}{2}u^{2}.
\end{array}
\end{equation}
Since the Killing and the dual metrics both satisfy the conditions we have
presented in Sec.4.1, they are also radiative in the sense of 
Lichn\'erowicz. 

For all of the subclasses we have investigated above we have found that
both the initial metric and its KS has only the $G_{44}$ component 
of the Einstein tensor surviving.  In a tensorial form this can be
expressed as
\begin{equation}
G_{\mu \nu}\,=\, \rho \, l_{\mu}l_{\nu},
\end{equation} 
where $\rho$ is the energy density, and its expression can be found 
straightforwardly, for each subclass.
Within the framework of Einstein's theory of relativity, this means that 
they describe pure radiative spacetimes~\cite{ksm}.

Once the arbitrary functions are specified then $k_{\mu \nu}$ can be found 
explicitly.  Finally, we would like to emphasize that, 
independent of the explicit forms of those arbitrary functions,
the dual metrics are also in the form of PN1P, satisfy LRC and 
are pure radiative. 

\section{Conclusion}
\noindent
In this paper, we have classified spacetimes with a field
of parallel null one-planes admitting non-degenerate Killing tensors.
In general, for an arbitrary metric, one cannot predict in advance that 
the Killing tensor equations
admit non-degenerate and non-trivial solutions, because there is not a well
defined technique to solve this problem.  For this purpose we have  
analyzed in detail equation (\ref{kte}) and looked for 
non-trivial Killing tensors that are of the same form as that of the initial 
metric.

The next step has been to evaluate the dual spacetimes associated with
those Killing metrics.
We have put some additional conditions on the metrics 
defining PN1P so that they describe radiative spacetimes; namely we have 
imposed Lichn\'erowicz radiation conditions.  
Furthermore, it can be seen by direct calculations that, to generate 
pure radiative spacetimes, it is sufficient to impose LRC.

We have found out that the dual spaces 
also satisfy the same properties as that of the initial ones,
endowing a physical significance to dual spaces as being pure 
radiative spacetimes.

Spacetimes with PN1P are under further investigation from a supersymmetric
point of view in connection with their Killing-Yano tensors~\cite{bb01}.

\section*{Acknowledgments}
\noindent
We would like to thank to M. Cahen and F. {\"O}ktem for stimulating 
discussions.

\end{document}